\documentclass[twocolumn]{aastex62}

\graphicspath{{./}{figures/}}

\received{\today}
\revised{\today}
\accepted{\today}
\submitjournal{ApJ}

\usepackage{lipsum}
\usepackage{calc} 

\usepackage{mathrsfs} 
\usepackage{amsmath}
\usepackage{needspace}

\usepackage{ragged2e}

\shorttitle{Testing Stellar Evolution with Asteroseismic Inversions}
\shortauthors{Bellinger et al.}

\defcitealias{1973A&A....23..325E}{Eggleton, Faulkner \& Flannery}

\newif\ifref
\reffalse
\definecolor{darkerred}{rgb}{0.25, 0, 0}
\newcommand{\mb}[1]{\ifref\textcolor{darkerred}{#1}\else #1\fi}

\newif\ifreff
\refftrue
\refffalse
\definecolor{darkred}{rgb}{0.7, 0, 0}
\newcommand{\mbb}[1]{\ifreff\textcolor{darkred}{#1}\else #1\fi}

\newif\ifrefff
\reffftrue
\reffffalse
\definecolor{darkred}{rgb}{0.7, 0, 0}
\newcommand{\mbbb}[1]{\ifrefff\textcolor{darkred}{#1}\else #1\fi}

\newif\ifreffff
\refffftrue
\refffffalse
\definecolor{darkred}{rgb}{0.7, 0, 0}
\newcommand{\mbbbb}[1]{\ifreffff\textcolor{darkred}{#1}\else #1\fi}

\newif\ifrefffff
\reffffftrue
\reffffffalse
\definecolor{darkred}{rgb}{0.7, 0, 0}
\newcommand{\mbbbbb}[1]{\ifrefffff\textcolor{darkred}{#1}\else #1\fi}

\begin{document}

\title{Testing stellar evolution with asteroseismic inversions of a main sequence star harboring a small convective core} 

\correspondingauthor{Earl P.~Bellinger}
\email{bellinger@phys.au.dk}

\author[0000-0003-4456-4863]{Earl P.~Bellinger} 
\altaffiliation{SAC Postdoctoral Fellow} 
\affil{Stellar Astrophysics Centre, 
Department of Physics and Astronomy, 
Aarhus University, 
Denmark} 

\author[0000-0002-6163-3472]{Sarbani Basu}
\affil{Department of Astronomy, Yale University, New Haven, CT, USA}

\author[0000-0002-1463-726X]{Saskia Hekker}
\affil{Max Planck Institute for Solar System Research, G\"ottingen, Germany}
\affil{Stellar Astrophysics Centre, Department of Physics and Astronomy, Aarhus University, Denmark}

\author[0000-0001-5137-0966]{J{\o}rgen Christensen-Dalsgaard} 
\affil{Stellar Astrophysics Centre, 
Department of Physics and Astronomy,  
Aarhus University, 
Denmark}

\begin{abstract}
The goal of stellar evolution theory is to predict the structure of stars throughout their lifetimes. 
Usually, these predictions can be assessed only indirectly, for example by comparing predicted and observed effective temperatures and luminosities. 
Thanks now to asteroseismology, which can reveal the internal structure of stars, it becomes possible to compare the predictions from stellar evolution theory to actual stellar structures. 
In this work, we present an inverse analysis of the oscillation data from the solar-type star KIC~6225718, which was observed by the \emph{Kepler} space observatory during its nominal mission. 
As its mass is about 20\% greater than solar, this star is predicted to transport energy by convection in its nuclear-burning core. 
We find significant differences between the predicted and actual structure of the star in the radiative interior near to the convective core. 
\mb{In particular, the predicted sound speed is higher than observed in the deep interior of the star, and too low at a fractional radius of 0.25 and beyond.} 
The cause of these discrepancies is unknown, and is not remedied by known physics in the form of convective overshooting or elemental diffusion. 
\end{abstract}


\section{Introduction} \label{sec:intro} 
Asteroseismology provides the unique opportunity to learn about the internal properties of stars through their global modes of oscillation. 
This is often achieved by fitting theoretical stellar models to observations of a star, and then assuming the best model to be a proxy for the star \citep[for an introduction to stellar modeling with asteroseismology, see][]{basuchaplin2017}. 
However, even the best theoretical models currently fail to match all of the asteroseismic signals that are observed in \mbbbb{stars exhibiting solar-like oscillations}, even after applying corrections for near-surface effects \citep[e.g.,][and references therein]{2014A&A...568A.123B}. 
This implies that the internal \mbb{structure} of models \mbbbb{of solar-like stars} are not exactly right, and motivates the need for another approach. 

The oscillations that are observed in \mbbbb{low-mass main sequence stars like the Sun} travel as sound waves through the stellar interior \citep[for an overview of asteroseismology, see][]{2010aste.book.....A}. 
As such, at each point in the interior these waves propagate at the local speed of sound, \mbb{the square of which} is proportional to the ratio of the pressure to density at that location in the stellar interior. 
When enough modes are observed, it becomes possible to map out the speed of sound at various points in the stellar interior, and thus infer the structure of the star. 
This analysis is referred to as a \emph{structure inversion}, as it is inverse to the forward problem of calculating the oscillation frequencies of a known stellar structure. 

Structure inversions are typically carried out by \mbb{inferring} the differences in internal structure between the star and a given model, called the reference model, through an inspection of the differences in their oscillation frequencies. 
This can be achieved by linearizing the perturbed stellar oscillation equations around the reference model \citep{Gough1985, 1991sia..book..519G}. 
When the reference model is the best-fitting stellar evolution model, a structure inversion leads directly to an evaluation of whether stellar evolution theory produces the correct stellar \mbb{structure} within the uncertainty of the inversion result. 

In the case of the Sun, for which there is rich helioseismic data, structure inversions have provided numerous constraints on the physics of the solar interior (see \citealt{2016LRSP...13....2B} for a review). 
For example, \citet{1993ApJ...403L..75C} compared the actual solar structure as inferred by seismic inversion to standard solar models which were calculated both with and without considering the effects of element diffusion. 
They found that the inclusion of diffusion \mb{deepened the convective envelope}, which ultimately resulted in a reduction of the discrepancy between theory and observation in the internal adiabatic sound speed profile by a factor of three, from a maximum difference of about 0.6\% down to about 0.2\%. 
Similarly, \citet{1997A&A...322L...5B} compared solar models that were constructed using different equations of state, and from this investigation they were able to rule out the \citetalias{1973A&A....23..325E} 
(EFF, \citeyear{1973A&A....23..325E}) EOS for solar matter. 
Although even today there remain highly significant discrepancies between the structure of solar models and the actual structure of the Sun, these analyses confirmed that solar models are nevertheless extremely accurate in an absolute sense. 

Thanks to the high-quality asteroseismic data collected by the NASA \emph{Kepler} space observatory \citep{2010Sci...327..977B}, we \mb{have the} first opportunity to perform similar tests of internal physics on other stars. 
There are however important differences between helioseismic and asteroseismic data, at least at the present time. 
Because the solar disk can be resolved, thousands of modes of high spherical degree ($\ell \lessapprox 250$) can be observed in the Sun. 
In stars, on the other hand, geometrical cancellation restricts observations to, at most, dozens of global modes ($\ell \leq 3$). 
Fortunately, nearly all of these modes traverse nearly all of the stellar interior, thereby providing constraints on the nature of the stellar core. 
The structure of stellar envelopes, on the other hand, are \mbb{insensitive} to asteroseismic \mbb{characterization}, at least at the present time. 
Another important difference is that we have accurate and precise independent knowledge of the fundamental solar parameters (age, mass, radius). 
For other stars, these must be estimated---usually also from the asteroseismic data---leading to large uncertainty in the results. 

Recently, \citet{2017ApJ...851...80B} introduced a technique to overcome these challenges by inverting an array of reference models spanning the uncertainties in the stellar mass and radius. 
Using this technique it was possible to resolve the radial structure of the cores in the two solar-like stars belonging to the 16~Cygni system. 
The uncertainties in the inversion result were large, however, making it difficult to assess whether there were important differences between the \mbb{structure} of the stars and the predictions of stellar evolution theory. 

In this paper, we take another approach, which was outlined in anticipation of first asteroseismic data by \citet{Basu2003}. 
Instead of determining the dimensional structure of a star, the precision of which is highly impacted by the uncertainties in the stellar mass and radius, we rather seek its dimensionless structure. 
The star we have selected for this analysis is KIC~6225718, also known in the literature as Saxo2, which is one of the best solar-type stars observed by \emph{Kepler} \citep{2017ApJ...835..172L}. 
\mbbbb{Like the Sun and 16~Cyg~A and B, it is a slow rotator, with a projected rotational velocity of {2.4 $\pm$ 0.5} km$/$s \citep{2013MNRAS.434.1422M}. 
Other observational constraints for this star are listed in Table~\ref{tab:fund-params}.}
Unlike the Sun and 16~Cyg~A and B, this star is just massive enough for stellar evolution theory to predict it to harbor a small convective core on the main sequence \citep[][]{2019A&A...622A.130B, 2019MNRAS.486.4612B}. 
Thus it constitutes an interesting testbed for furthering our understanding of stellar physics. 

\begin{table}
    \centering
    \caption{\mbbbb{Observed Parameters of KIC~6225718} \label{tab:fund-params}}
        \begin{tabular}{ccc}\hline
            Parameter & Value & Unit \\\hline\hline
            $T_{\text{eff}}$ & $6313 \pm 77$ & K \\
            $\text{[Fe/H]}$ & $-0.07 \pm 0.10$ & \\
            $\nu_{\max}$ & $2369 \pm 24$ & $\mu$Hz \\
            $\Delta\nu$ & $105.754 \pm 0.096$ & $\mu$Hz \\\hline
        \end{tabular}
    \vspace*{-2mm}\justify
    \emph{Notes}. Rows contain the effective temperature, metallicity, frequency at maximum oscillation power, and large frequency separation. Values adopted from \citet{2017ApJ...835..172L}.
\end{table}

\section{Evolutionary modelling} 
The first step of the analysis is to obtain a suitable reference model for KIC~6225718. 
We used the \emph{Stellar Parameters in an Instant} \citep[SPI,][]{2016ApJ...830...31B, 2017ApJ...839..116A} method to estimate the fundamental parameters of KIC~6225718 and subsequently \mbb{obtained} a model according to those parameters. 
Briefly, this method uses machine learning with theoretical stellar models to determine which parameters (mass, age, initial chemical composition, mixing length parameter, etc.) are most consistent with the observations of a given star. 
As this method requires a grid of models, we used \emph{Modules for Experiments in Stellar Astrophysics} \citep[MESA r10108,][]{2011ApJS..192....3P,2013ApJS..208....4P,2015ApJS..220...15P,2018ApJS..234...34P, 2019arXiv190301426P} for the evolution calculations. 
We calculated 1024 tracks with initial conditions varied quasi-randomly using a Sobol generation scheme \citep[see Appendix~B of][]{2016ApJ...830...31B}. 
The tracks were varied in mass (${0.7 \leq M/\text{M}_\odot \leq 3}$), initial helium abundance (${0.22 \leq Y_0 \leq 0.34}$), initial metallicity (${0.0001 \leq Z_0 \leq 0.04}$ \mbb{on a uniform logarithmic grid}), and mixing length parameter ($1 \leq \alpha_{MLT} \leq 3$).
Diffusion and overshoot are not considered at this stage of the analysis, but will be considered later. 
The remaining aspects of the models are the same as described by \citet[][]{2019A&A...622A.130B}. 
\mbbb{We used the mean-shift algorithm \citep{fukunaga1975estimation} to find the mode of the joint posterior distribution of the estimated parameters (age, mass, chemical composition, and mixing length parameter, \mbbbb{tabulated in Table~\ref{tab:ref-mod}}), and used those parameters to compute the reference model.} 

\begin{table}
    \centering
    \caption{\mbbbbb{Estimated Parameters of KIC~6225718} \label{tab:ref-mod}}
        \begin{tabular}{ccccc}\hline
            Parameter & Estimate & Ref.\ Mod. & Solar & Unit \\\hline\hline
            Age & $3.09\pm 0.63$ & 2.803 & 4.572 & Gyr \\
            Mass & $1.189 \pm 0.053$ & 1.20 & 1 & M$_\odot$ \\
            Radius & $1.25 \pm 0.024$ & 1.24 & 1 & R$_\odot$  \\
            $X_c$ & $0.287\pm 0.053$ & 0.30 & 0.34 & -- \\
            $X_0$ & $0.7080 \pm 0.0059$ & 0.71 & 0.70 & -- \\
            $Z_0$ & $0.0202 \pm 0.0031$ & 0.021 & 0.019 & -- \\
            $\alpha_{\text{MLT}}$ & $2.04\pm 0.14$ & 2.05 & 1.84 & -- \\\hline
        \end{tabular}
    \vspace*{-2mm}\justify
    \emph{Notes}. The quantity $X_c$ refers to the fractional hydrogen abundance in the core of the model, which is a proxy for its main-sequence age. The quantities $X_0$ and $Z_0$ refer to the initial hydrogen abundance and initial metallicity, respectively. The quantity $\alpha_{\text{MLT}}$ refers to the mixing length parameter. The values of the reference model as well as those for a solar-calibrated model are provided. 
\end{table}

\begin{figure}
    \centering
    \includegraphics[width=\linewidth, trim={0.05in 0 0.15in 0}, clip]{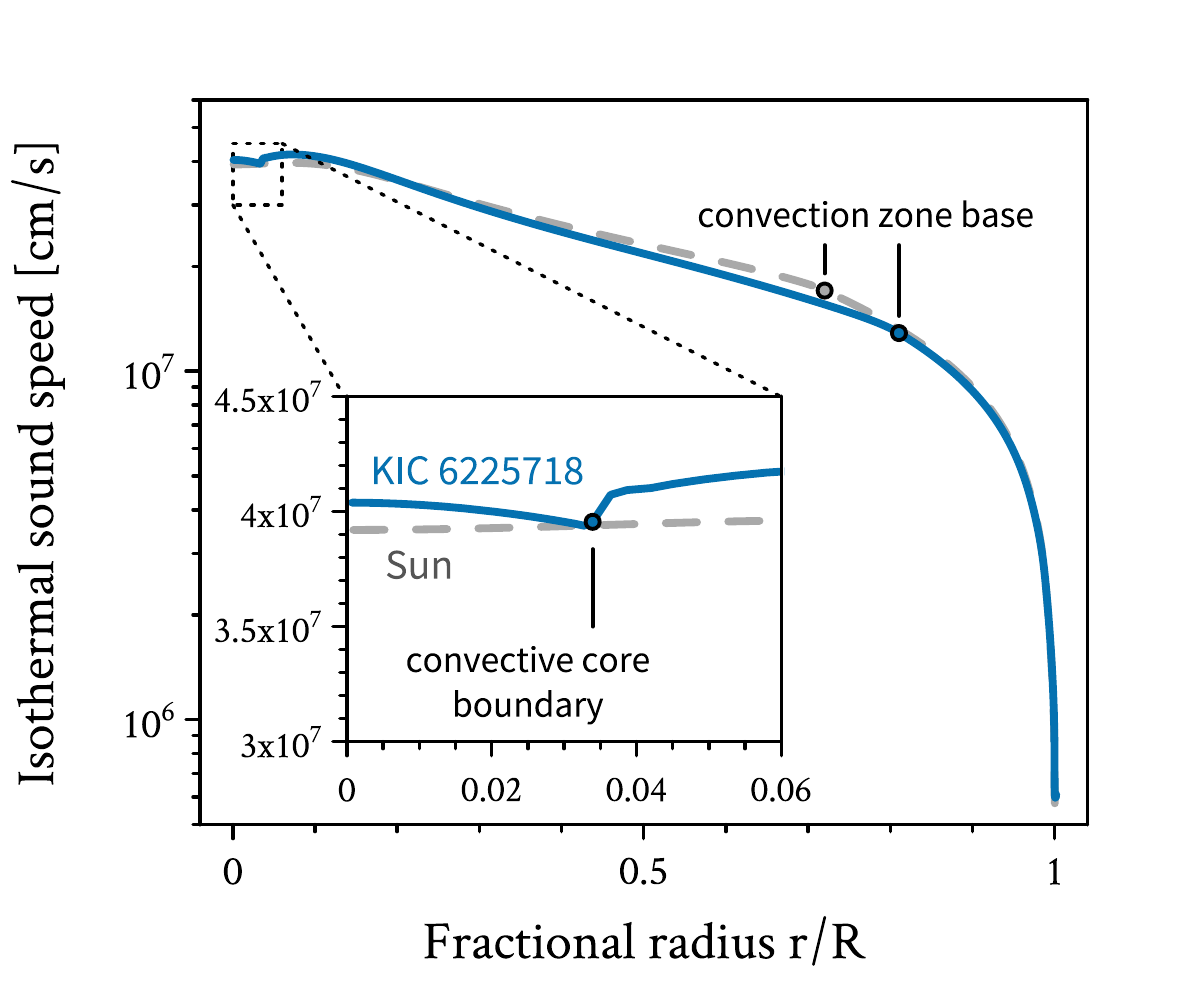}
    \caption{The acoustic structure \mbbb{$\sqrt u$} of the best-fitting model for KIC~6225718 as a function of the fractional radius ${r/R}$, where $r$ is the distance from the stellar center and $R$ is the total stellar radius. 
    A solar model is shown for reference. 
    The inset diagram shows a zoom-in on the deep stellar interior, where the signature of a convective core is apparent. 
    Convective boundaries are indicated. }
    \label{fig:structure}
\end{figure}

\begin{figure*}
    \centering
    \includegraphics[width=0.5\linewidth]{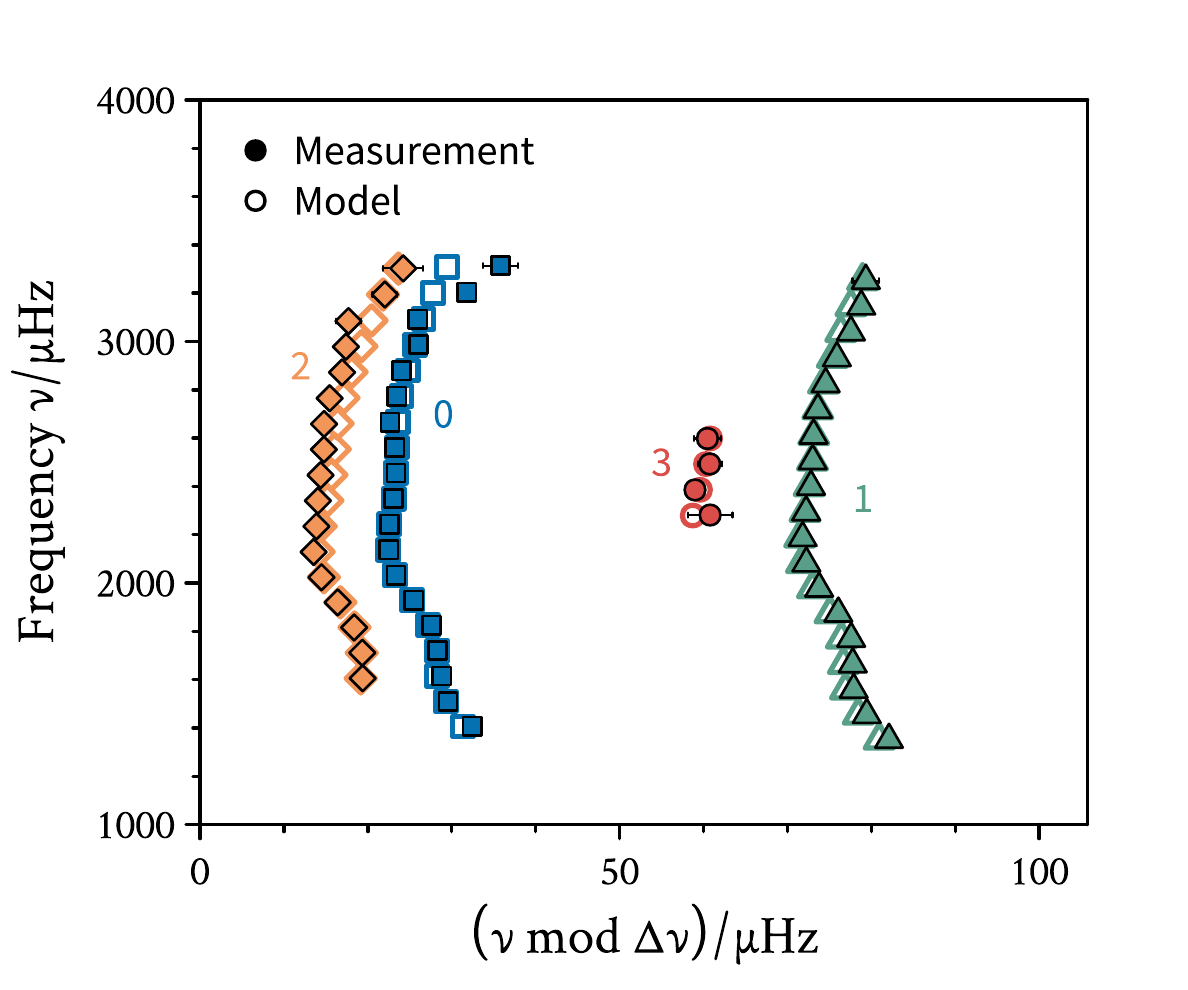}%
    \includegraphics[width=0.5\linewidth]{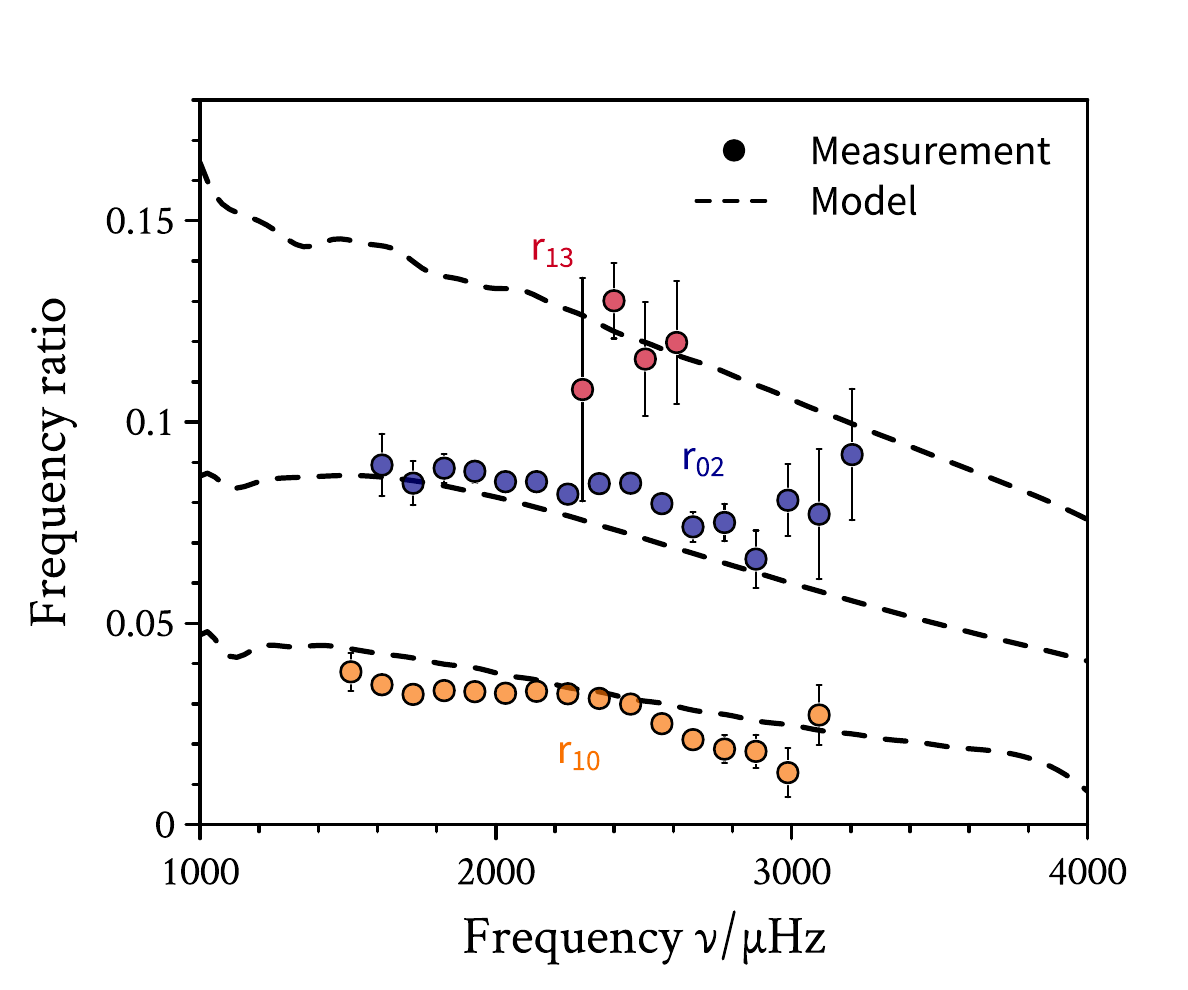}%
    \caption{Comparisons of predicted and measured seismic \mbb{properties}. 
    \textsc{Left Panel}. \'{E}chelle diagram comparing the observed frequencies of KIC~6225718 to the frequencies of the best-fitting stellar model, \mbbb{which have been corrected for surface effects}. 
    The spherical degrees of the modes are labelled. 
    The uncertainties of the measurements are indicated, the majority of which are too small to be seen. 
    \textsc{Right Panel}. A comparison of observed and (interpolated) theoretical frequency ratios. \\[0.2\baselineskip] } 
    \label{fig:echelle}
\end{figure*}

\mb{Figure~\ref{fig:structure} shows the structure of the reference model. In particular, it shows the internal profile of the isothermal speed of sound} $\sqrt u$, defined as $u= P/\rho$,
with $P$ being pressure and $\rho$ being density. 
Figure~\ref{fig:echelle} shows an asteroseismic comparison of this model to the observations of KIC~6225718. 
Two ways of dealing with the surface term are shown: one by applying the \citet{2014A&A...568A.123B} two-term correction, and another by inspecting the asteroseismic frequency ratios $r_{02}$, $r_{13}$, and $r_{10}$ \citep[for definitions, see, e.g.,][]{2003A&A...411..215R, 2013A&A...560A...2R, 2005A&A...434..665R, 2018arXiv180807556R}. 
Significant differences are apparent in 17 of the modes, with the differences in some modes exceeding 8$\sigma$. 
In both cases it is clear that the stellar evolution model does not pulsate the same way as does the star. 
Furthermore, these differences are caused by differences in the internal structure of the star, and not by surface effects. 
The task is then to find where the internal structure differs.

\needspace{3\baselineskip}
\section{Asteroseismic inversion} 
We now seek to obtain asteroseismic measurements of the internal sound speed profile so that we may compare the structure of the star with the structure that is predicted from stellar evolution theory. 
In order to achieve this, we perturb the equations of stellar oscillation and linearize them around the reference model in order to determine how changes to the stellar structure result in changes to the mode frequencies \citep[e.g.,][]{basuchaplin2017}. 
For each mode of oscillation we obtain an equation which relates the difference in the frequency of the mode between the star and the reference model to the differences in their internal structure: \mbb{
\begin{equation}
    \frac{\delta \nu}{\nu}
    =
    \int
    K^{(u',Y)}
    \,\frac{\delta u'}{u'}
    \,\text{d}x
    +
    \int
    K^{(Y,u')}
    \,\delta Y
    \,\text{d}x.
\end{equation}
}Here $\nu$ is the frequency of the mode, $\delta\nu$ is the difference in $\nu$ between the model and the star, $\delta Y$ is the difference in the fractional abundance of helium, and $\delta u'$ is the difference in the dimensionless squared isothermal sound speed $u'=uR/M$, with $M$ being the total stellar mass \citep{Basu2003}. 
The kernels, denoted $K$, are computed numerically based on the structure of the reference model \citep[e.g.,][]{1991sia..book..519G, 1999JCoAM.109....1K}. 
These functions quantify how a perturbation to $u'$ or $Y$ at any point in the stellar interior will result in a perturbation to $\nu$. 
Examples of these kernels for various oscillation modes were presented by \citet{2017ApJ...851...80B}. 

\mbb{The reason for having two kernels rather than one is that the frequencies of oscillation depend on dynamical variables (pressure, density, and gravity) as well as the adiabatic compressibility $\Gamma_1 = (\partial \ln P / \partial \ln \rho)_{\text{ad}}$. 
Whereas each dynamical variable is directly expressible in terms of the others and thus can be used to eliminate them, $\Gamma_1$ depends on the equation of state and thus on the composition (and hence the helium abundance) of the stellar plasma.} 
Fortunately, the helium kernel has negligible amplitude throughout the majority of the stellar interior, which effectively isolates differences in the frequencies of the mode to differences in the sound speed. 
Although deriving the helium kernel comes at the cost of assuming an equation of state, the systematic error caused by this assumption is likely to be much smaller than the uncertainty on the inversion result \citep{1997A&A...322L...5B}. 

As 59 oscillation modes have been detected by \emph{Kepler} in KIC~6225718, we have 59 such equations to work with. 
This forms the set of equations \mbbb{that we use} to determine the differences in $u'$ between the star and the best-fitting stellar model. 

In order to combine the measurements of mode frequencies to produce a measurement of the internal structure, we use the inversion technique known as Subtractive Optimally Localized Averages \citep[SOLA,][]{1992A&A...262L..33P, 1994A&A...281..231P}. 
This method works by combining the $u'$ kernels in such a way that their combination, known as the averaging kernel (${\mathscr{K} = \sum_i c_i K_i^{(u',Y)}}$, where $\mathbf c$ are the coefficients of the linear combination and $i$ refers to the $i$th observed mode of oscillation), only has amplitude \mbb{near} one location, called the target radius ($r_0$). 
A well-localized averaging kernel represents an instrument for inferring differences in $u'(r_0)$ between the star and the model. 
We achieve this by optimizing $\mathbf c$ such that the resulting averaging kernel best resembles a function of our choosing, called the target kernel, which we choose to be a Gaussian peaked at $r_0$ modified such that it goes smoothly to zero at the core \citep[e.g.,][]{basuchaplin2017}. 
Provided the averaging kernel is well-localized at the target radius and integrates to unity, and furthermore if that same combination of $Y$ kernels (called the cross-term kernel, ${\mathscr{C}= \sum_i c_i K_i^{(Y,u')}}$) has negligible amplitude everywhere, then an estimate of the structure of the star can be obtained by applying that same linear combination to the relative differences in the mode frequencies. 
In other words, \mbb{
\begin{equation}
    \sum_i c_i\, \frac{\delta \nu_i}{\nu_i}
    =
    \left\langle \frac{\delta u'}{u'} \right\rangle (r_0) 
    \simeq 
    \int \mathscr{K}\, \frac{\delta u'}{u'} \, \text{d}x 
\end{equation}
where the angled brackets represent a weighted average, with the averaging kernel being the weighting function.
Surface effects are suppressed by enforcing 
\begin{equation}
    \sum_i c_i \, F_\text{surf}(\nu_i) = 0
\end{equation}
where we adopt the \citet{2014A&A...568A.123B} two-term correction for the surface term $F_\text{surf}$. 
The uncertainty in the solution is 
\begin{equation}
    e^2 = \sum_i c_i^2 \sigma_i^2
\end{equation}
where $\sigma_i$ is the observational uncertainty of the frequency of mode $i$.
}Thus when optimizing the coefficients $\mathbf c$ subject to the aforementioned constraints, we tune free parameters which balance the agreement of the averaging kernel with the target kernel against the resulting amplitude of the cross-term kernel as well as the magnification of the uncertainty in the results \citep[e.g.,][]{RabelloSoares1999}. 

\begin{figure*} 
    \centering%
        \includegraphics[width=0.5\linewidth]{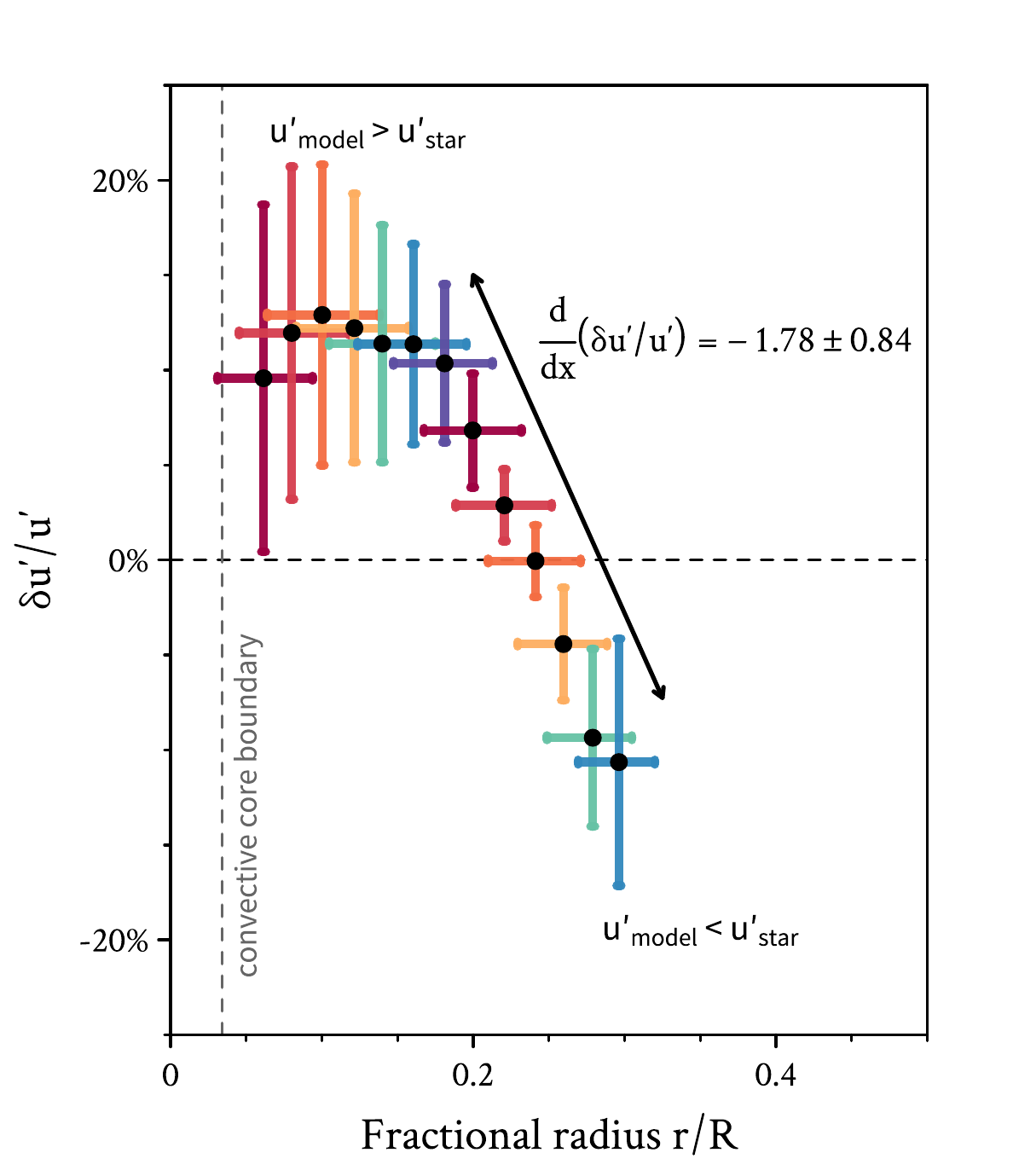}%
    \begin{minipage}[b]{0.5\textwidth}%
        \includegraphics[width=\linewidth, trim={0 0.57in 0 0}, clip]{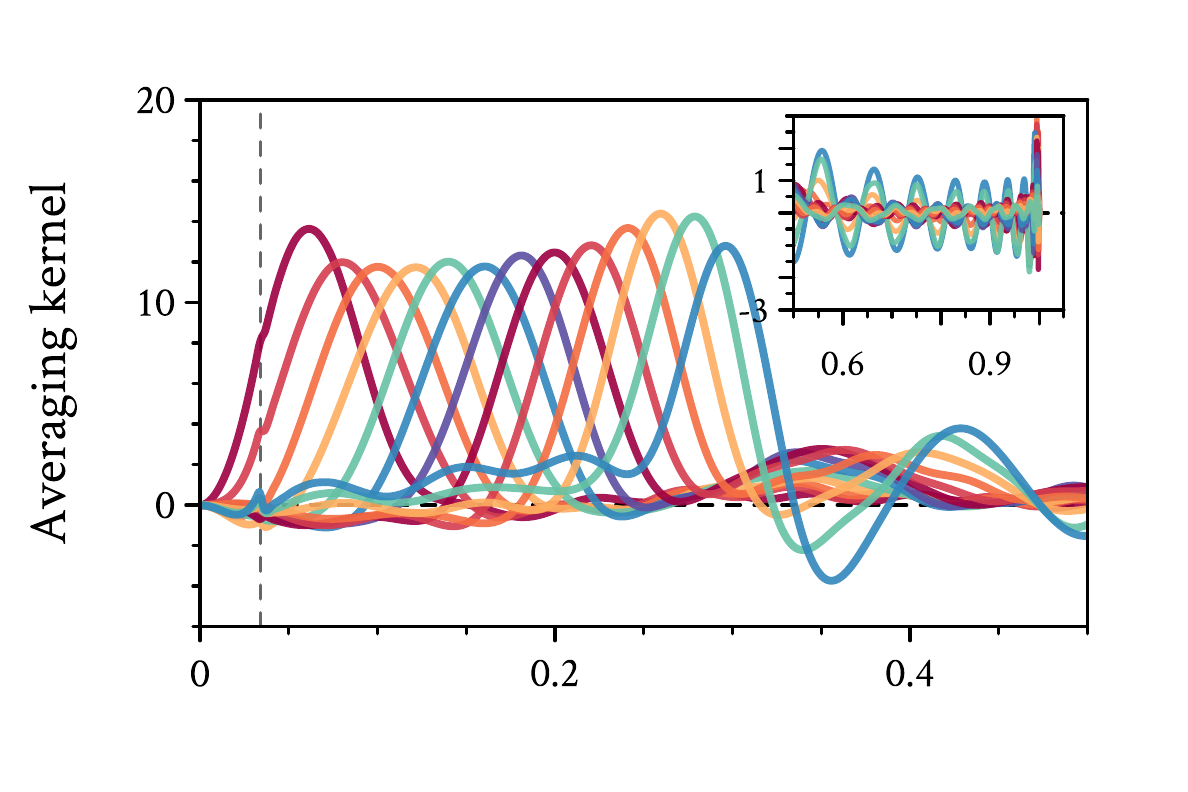}\\%
        \includegraphics[width=\linewidth, trim={0 0 0 0.27in}, clip]{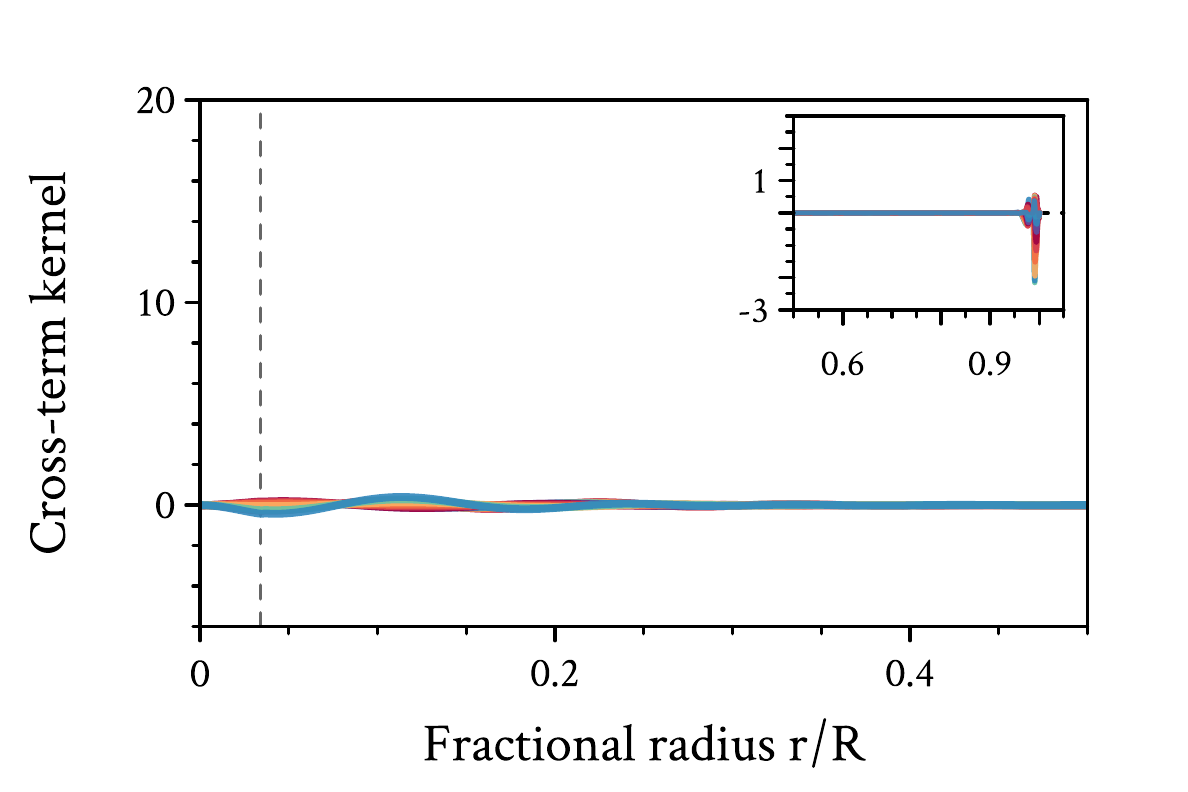}%
    \end{minipage}\\%
    \caption{Result of the asteroseismic structure inversion. \textsc{Left Panel}. 
    Relative difference between the dimensionless squared isothermal sound speed in the star and the best-fitting stellar model, in the sense of $(\text{model}-\text{star})/\text{model}$. 
    A strong gradient can be seen, the slope of which is indicated. 
    The horizontal uncertainties correspond to the width of the averaging kernel. 
    The colors are for ease of identifying the corresponding averaging kernel. 
    \textsc{Top Right Panel}. Averaging kernels colored by target radius, corresponding to the inversion results in the left panel. 
    The convective core boundary at ${x = r/R\simeq 0.035}$ is visible as a small bump in the averaging kernels and is indicated with a dashed gray line. 
    The inset figure shows the outer layers of the star. 
    Note the differences in scale. 
    \textsc{Bottom Right Panel}. Cross-term kernels for the inversion. 
    For ease of comparison, the scales are set to match those of the averaging kernels. 
    The inset again shows the behavior in the outer layers. 
    The convective core boundary is indicated. 
    \\[0.2\baselineskip]
    \label{fig:inv-results} }  
\end{figure*}

The results of the asteroseismic inversion are shown in Figure~\ref{fig:inv-results}. 
Well-localized averaging kernels were able to be formed at target radii ${0.06 - 0.3 R}$, probing the radiative interior of the star just beyond the convective core boundary (${0.034 R}$). 
The cross-term kernels have negligible amplitude everywhere, and the contributions at the surface are small. 

\begin{figure}
    \centering
    \includegraphics[width=\linewidth, trim={0.05in 0 0.15in 0}, clip]{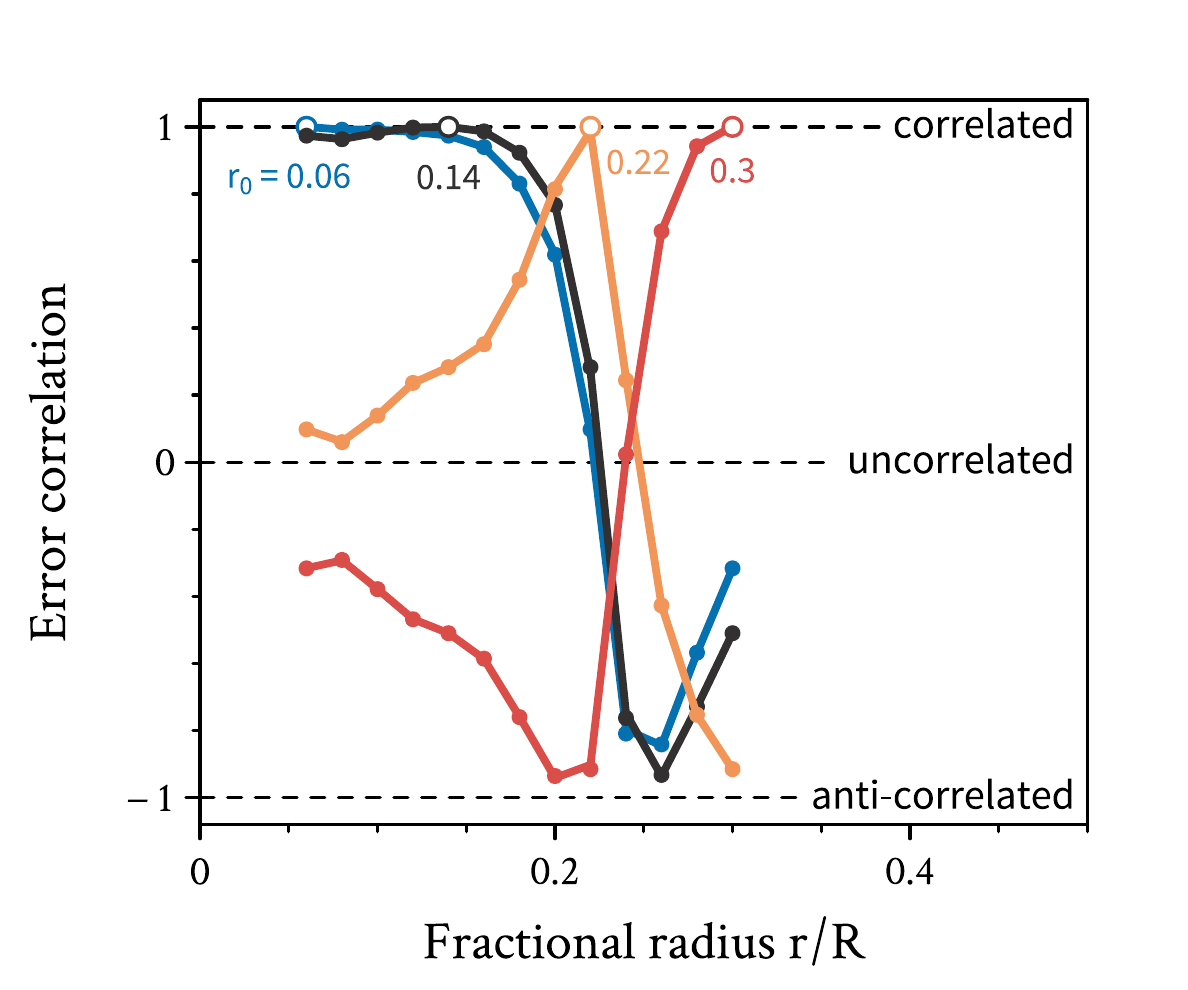}
    \caption{Error correlations in the inversion results for four target radii \citep[see Equation~10.38 in][]{basuchaplin2017}. 
    Each curve shows a comparison of the error correlation between the result at that target radius (indicated with an open circle) and all other target radii (closed circles).  } 
    \label{fig:err_corr}
\end{figure}

Substantial differences between the structure of the star and the structure of the model as well as a large gradient are apparent in the inversion result. 
As the results of the inversion at different target radii are correlated (see Figure~\ref{fig:err_corr}) since they are each estimated using the same data, it is not trivial to assess the significance of this slope. 
In order to account for these correlated uncertainties, we performed 10\,000 Monte Carlo realizations of the uncertainty in the frequencies, and obtained $\text{d}/\text{d}x(\delta u'/u') = -1.78 \pm 0.84$ as indicated in the figure, where ${x= r/R}$. 
Thus the slope differs significantly from zero at a level of more than 2$\sigma$. 
This indicates that there are significant differences between the structure of the star and the structure of the best-fitting stellar model.

\section{\mb{Alternative} models} 
Having found that there are differences between the star and the stellar model, we now seek to understand the source of these differences. 
This can be achieved through a comparison of the reference model and models which have been constructed differently. 
We consider four such possibilities: the effect of the radius-to-mass ratio of the reference model, the effect of the input physics (diffusion and overshoot), the effect of the evolution code and fitting technique, and the effect of the surface term correction. 

\subsection{Effect of mass and radius} 
The best-fitting model of KIC~6225718 was selected from the mode of the joint posterior distribution. 
Yet this is not the only model consistent with the observations; other models with a higher or lower mass or radius are not ruled out. 
Figure~\ref{fig:alternative-models} shows a comparison of the $u'$ structure between the reference model and two models which differ by $\pm1\sigma$ (${0.03~\text{R}_\odot/\text{M}_\odot}$) in the radius-to-mass ratio.
The resultant differences are small, and thus we can conclude from this comparison that differences in the global parameters of the model are unlikely to be the cause of the discrepancies seen in the inversion result. 
We furthermore performed the inversion with these models as reference and obtained the same result (not pictured). 

\subsection{Effect of input physics}%
In order to assess whether a different set of input physics would resolve the discrepancies, we computed three more grids of models which include the effects of (I) convective core overshoot, (II) element diffusion, and (III) both of these. 

\mbbbb{We used the classical step formulation of convective core overshooting with an overshooting parameter ${\alpha_{\text{ov}}=0.2}$. 
As is default in MESA, overshoot is treated with a radiative temperature gradient. 
We do not consider the effects of convective envelope undershooting.} 

Since gravitational settling in higher-mass models leads to the unobserved consequence of zero-metallicity atmospheres \mbbbb{unless it is resisted by mass loss or radiative levitation \citep[e.g.,][]{2018A&A...618A..10D}, the computation of which is computationally expensive}, we tapered off diffusion for evolutionary tracks with ${M> 1.25\;\text{M}_{\odot}}$ according to the equation given by \citet{2017EPJWC.16005005V}.
\mbbbb{Despite the unphysical nature of this prescription, the star under investigation here is of a low enough mass that the diffusion-tapered models are irrelevant.}
\mbbbbb{The neglect of radiative levitation should not make a difference to the result as its main effects are in the outer layers of the star.} 
The remaining aspects of the models were unchanged. 
We then used SPI to find the best-fitting model for each of these grids. 

Figure~\ref{fig:alternative-models} compares the $u'$ profiles of these models with the reference model. 
It can be seen that the differences caused by a change of input physics do not produce a signal of the same magnitude seen in the inversion result. 
We again inverted using these models as reference and again obtained essentially the same result. 
\mbbbb{Furthermore, the mode frequencies of these models have even larger deviations with respect to the observations than the reference model.} 
Thus this is an unlikely culprit.

\needspace{3\baselineskip}
\subsection{Effect of evolution code \& fitting technique} 
\begin{figure*}
    \centering
    \newlength\graphicwidth
    \setlength\graphicwidth{0.37\linewidth}
    \def\mygraphic{\includegraphics[width=\graphicwidth, trim={0 0 0.29in 0}, clip, keepaspectratio]{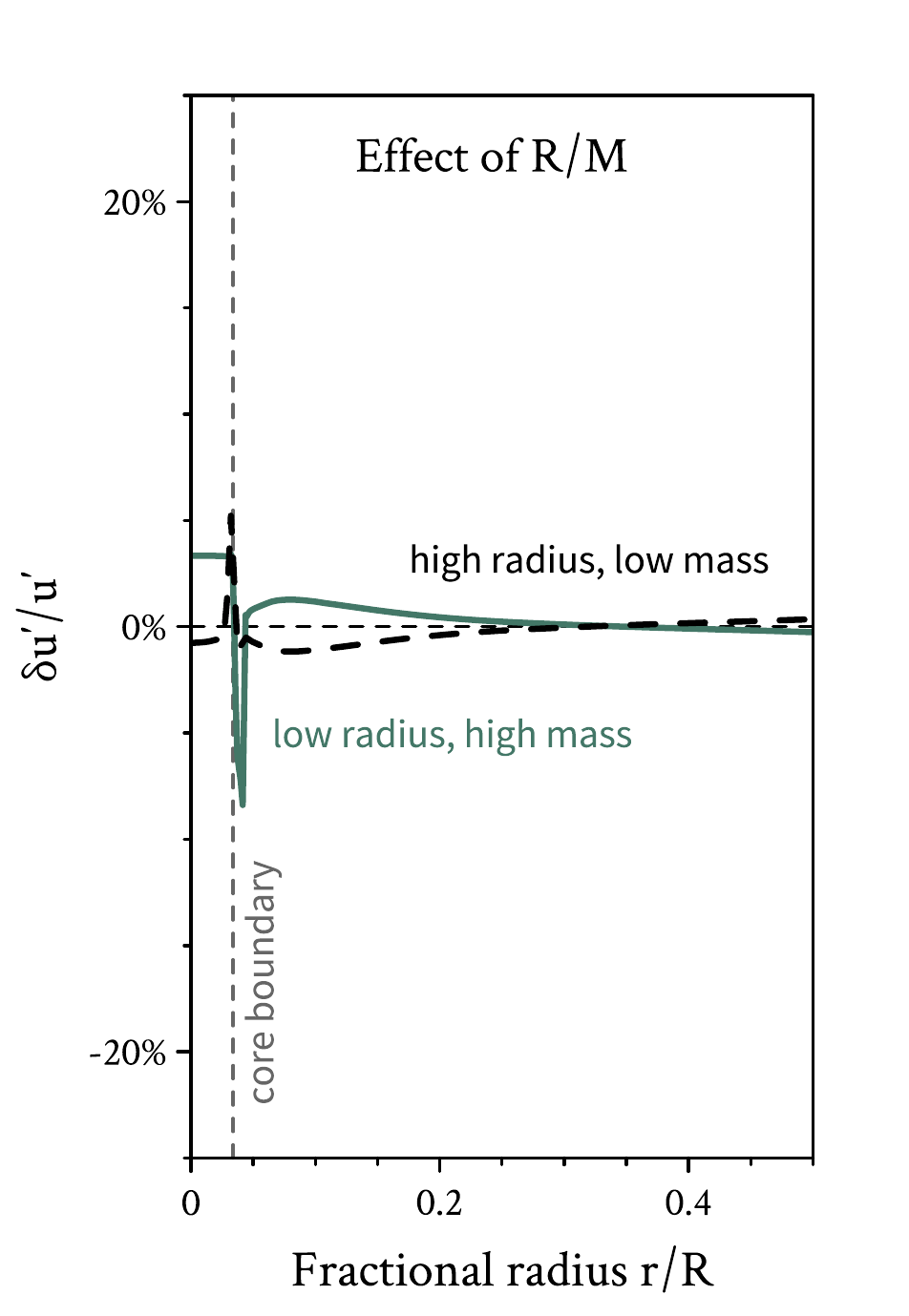}}%
    \newlength\graphicheight
    \setlength\graphicheight{\heightof{\mygraphic}}
    \mygraphic{}\hfill%
    \includegraphics[width=\graphicwidth, height=\graphicheight, trim={0.7in 0 0.29in 0}, clip, keepaspectratio]{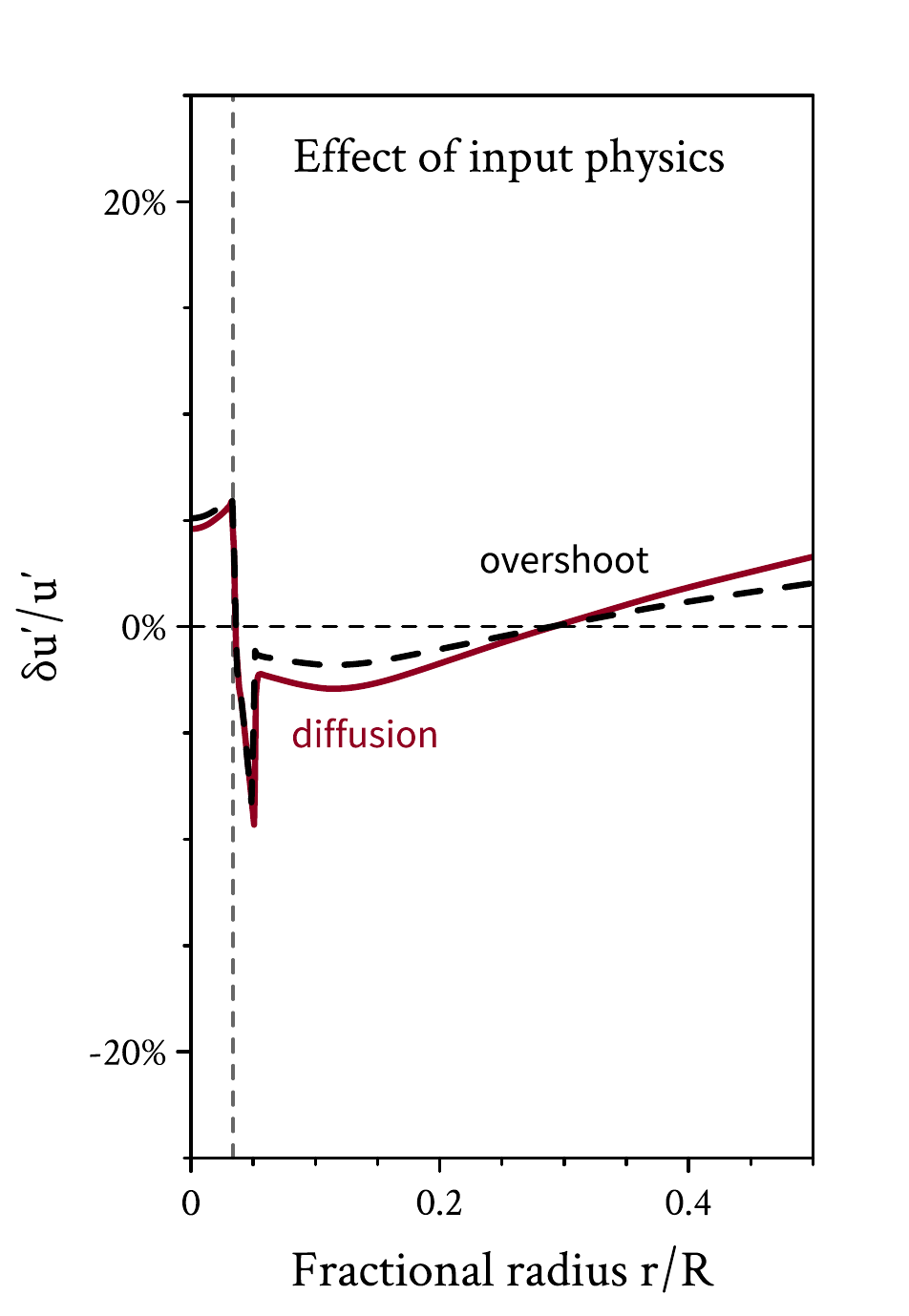}\hfill%
    \includegraphics[width=\graphicwidth, height=\graphicheight, trim={0.7in 0 0 0}, clip, keepaspectratio]{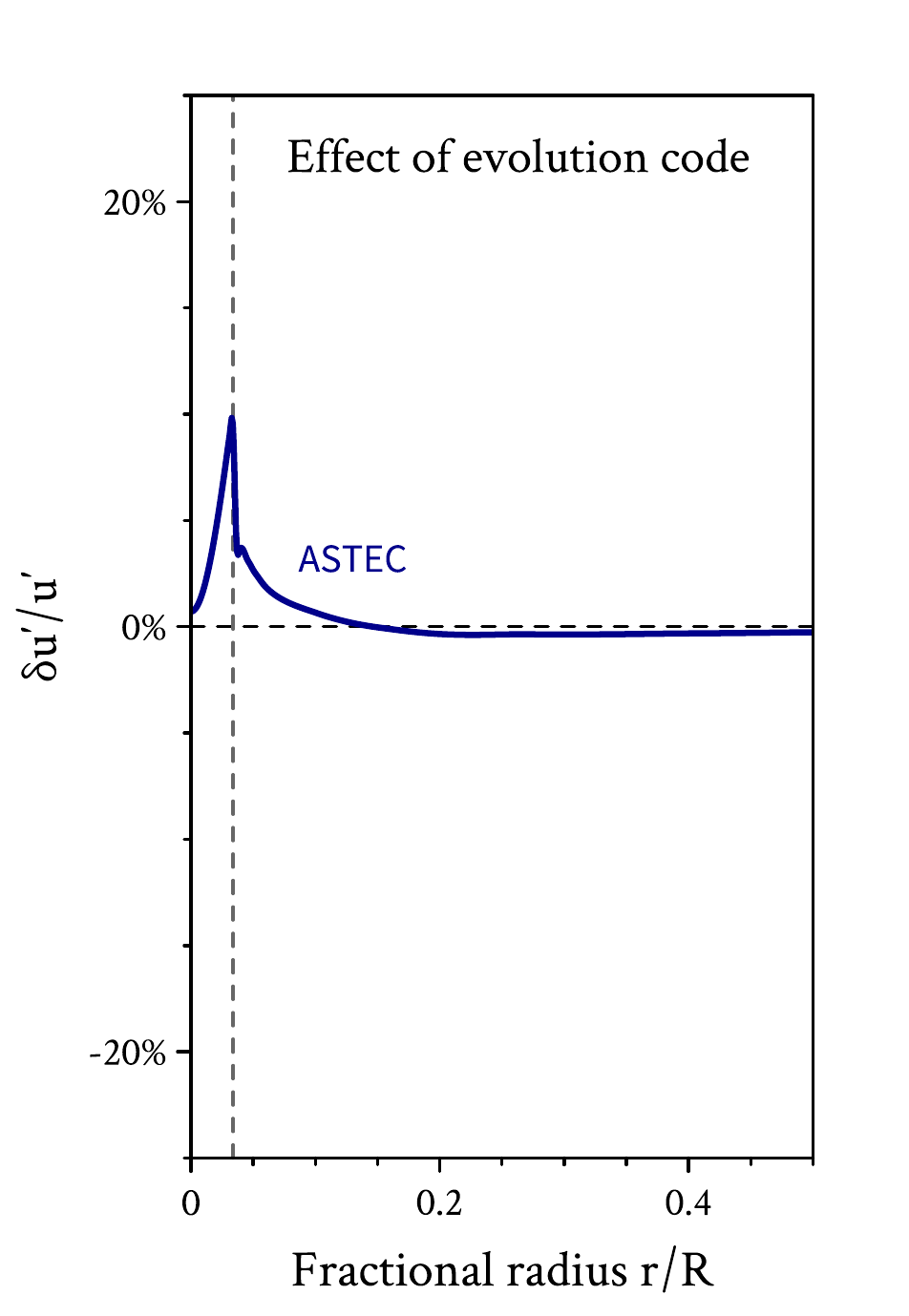}
    \caption{Relative difference in dimensionless sound speed between the original model and different models of KIC~6225718. For the purposes of comparison, the axis ranges have been made the same as those in Figure~\ref{fig:inv-results}. In each case, the differences are smaller than the differences seen between the star and the model. The boundary of the convective core in the original model is indicated. 
    \textsc{Left Panel}. Differences between the original model and models \mbb{which are higher and lower in their radius-to-mass ratio}. 
    \textsc{Middle Panel.} Differences between the original model and models of differing input physics. A model with both diffusion and overshoot (not pictured) behaves approximately like an average of the two individually. 
    \textsc{Right Panel}. Differences between the original MESA model and a model made using ASTEC. \label{fig:alternative-models} }
\end{figure*}

Figure~\ref{fig:alternative-models} compares the structure of the MESA/SPI model with the best-fitting model obtained using the Aarhus Stellar Evolution Code \citep[ASTEC,][]{2008Ap&SS.316...13C, 2017ApJ...835..173S}. 
The likelihood-weighted properties from ASTEC give a mass, radius, and age that are all within $1\sigma$ of the MESA/SPI estimates. 
The radius-to-mass ratio of the best-fitting ASTEC model is approximately the same as that of the best-fitting MESA model. 
We again find that an inversion using the ASTEC model as reference produces the same result as with the MESA model. 

\needspace{3\baselineskip}
\subsection{Effect of surface term correction}
\begin{figure*} 
    \centering 
    \includegraphics[width=0.5\linewidth]{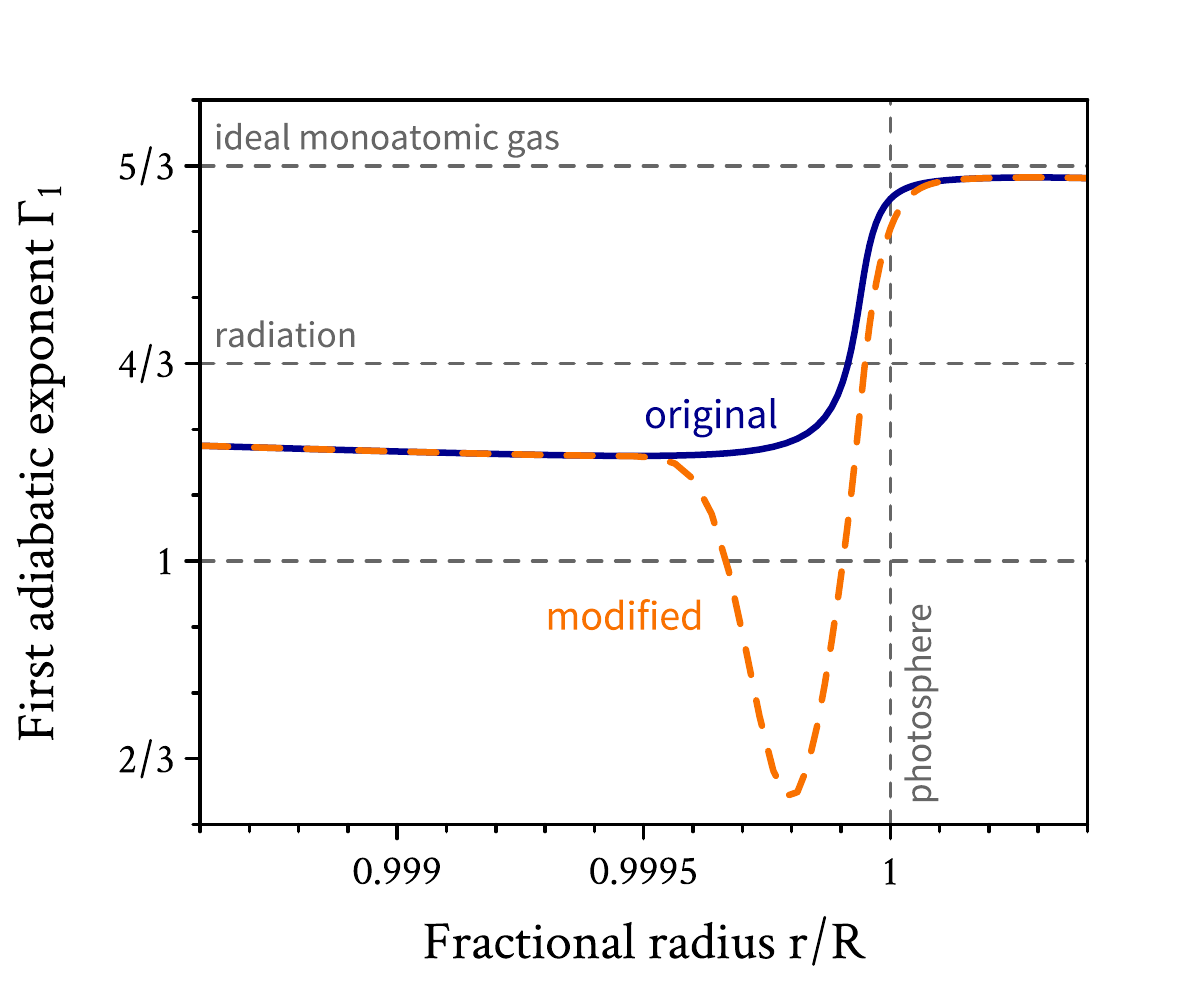}%
    \includegraphics[width=0.5\linewidth]{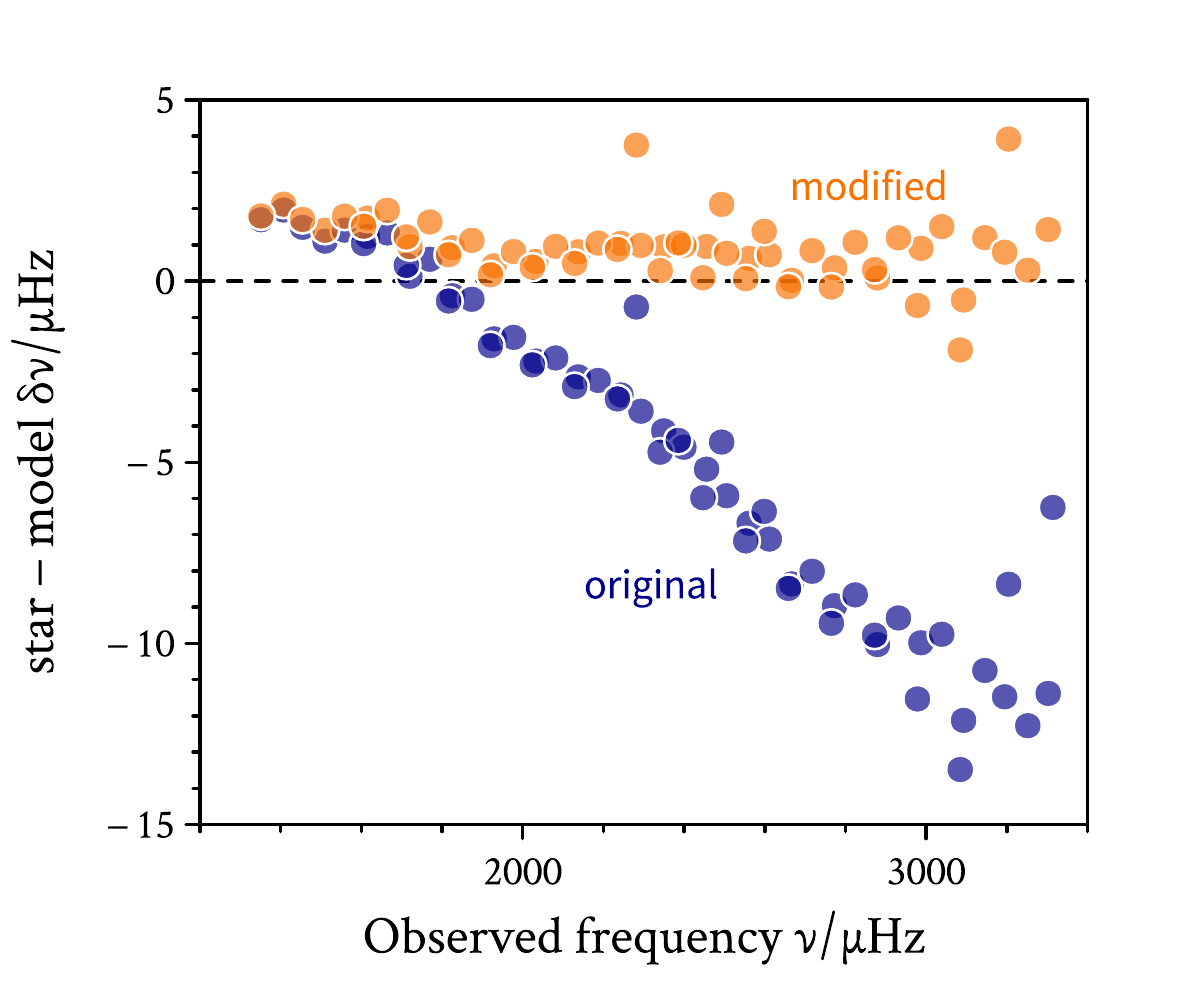}
    \caption{\textsc{Left Panel}. A comparison of the near-surface thermal structure of the original model and the model which has been modified to remove surface effects. \textsc{Right Panel}. A comparison of the mode frequencies of between the star and the two models shown on the left. \label{fig:gamma1} }  
\end{figure*}

As a final test, we consider a model with a surface structure artificially modified in a way such that it does not show a systematic error between the observed and theoretical model frequencies (i.e., no surface effect). 
We achieve this by changing \mbb{the adiabatic compressibility} in the near surface layers. 
Figure~\ref{fig:gamma1} shows $\Gamma_1$ in the modified model and a comparison of its oscillation mode frequencies, confirming that the (non-physical) modification to the near-surface $\Gamma_1$ profile has eliminated surface effects. 
An inversion with this modified model, however, again yields the same result. 
As the amplitude of the averaging and cross-term kernels are small near the surface anyway (Figure~\ref{fig:inv-results}), this result was expected. 

\needspace{3\baselineskip}
\section{Conclusions}
We have presented an asteroseismic investigation of the \emph{Kepler} target KIC~6225718. 
After obtaining a best-fitting stellar model of this star, we found that the asteroseismic \mbb{properties} of the star differ from those of the model, even after applying surface term corrections to the theoretical mode frequencies. 
This implies that the internal structure which is predicted by stellar evolution theory is not correct. 
We then performed an inverse analysis of the oscillation frequencies to determine the dimensionless internal structure of the star. 
Comparing this structure of the star with the structure of the model, we found there are significant differences in the structure of the radiative interior near to the convective core boundary. 
We investigated whether these differences could stem from having an incorrect mass or radius in the model, incorrect input physics, incorrect treatment of the surface term, or from the choice of stellar evolution code. 
In each case however we found that the differences imposed by such an effect are much smaller than the differences that we found between the star and the model. 
Thus the cause of these discrepancies remains a mystery. 

\mbbb{As the speed of sound is inversely related to the mean molecular weight, these results may imply that important internal mixing processes are missing from stellar evolution calculations.} 
\mbbbbb{While previous studies have indicated the need for additional mixing at the convective core boundary \citep[e.g.,][]{2016A&A...589A..93D}, we have found that convective core overshooting is insufficient to explain the internal structure of KIC~622517, and that additional processes are likely to be at work. }

\acknowledgments 
We thank the anonymous referee for useful comments. 
Funding for the Stellar Astrophysics Centre is provided by The Danish National Research Foundation (Grant agreement no.: DNRF106). 
S.H.\ acknowledges funding from the European Research Council under the European Community's Seventh Framework Programme (FP7/2007-2013) / ERC grant agreement no 338251 (StellarAges). 
S.B.\ acknowledges partial support from NSF grant AST-1514676 and NASA grant NNX13AE70G.

\software{ASTEC \citep{2008Ap&SS.316...13C}, ADIPLS \citep{2008Ap&SS.316..113C}, MESA \citep{2011ApJS..192....3P,2013ApJS..208....4P,2015ApJS..220...15P,2018ApJS..234...34P, 2019arXiv190301426P}, GYRE \citep{2013MNRAS.435.3406T, 2018MNRAS.475..879T}, R \citep{r}, magicaxis \citep{magicaxis}, scikit-learn \citep{scikit-learn}}


\clearpage
\bibliographystyle{aasjournal.bst}
\bibliography{Bellinger}

\end{document}